\documentclass[aps,prd,email,preprint,showkeys,showpacs,preprintnumbers,amsmath,amssymb,nofootinbib]{revtex4}

\usepackage{amsmath}
\usepackage{latexsym}
\def\[{\left\lbrack}
\def\]{\right\rbrack}

\def\({\left(}
\def\){\right)}

\def\ni{\noindent}
\newcommand{\be}{\begin{equation}}
\newcommand{\ee}{\end{equation}}
\newcommand{\ea}{\end{eqnarray}}
\newcommand{\ba}{\begin{eqnarray}}

\begin{document}

\title{\Large{Rotating traversable wormholes in a noncommutative space and the energy conditions}}

\author{Everton M. C. Abreu$^{a,b}$}
\email{evertonabreu@ufrrj.br}
\author{N\'elio Sasaki$^b$}
\email{sasaki@fisica.ufjf.br}

\affiliation{${}^{a}$Grupo de F\' isica Te\'orica e Matem\'atica F\' isica, Departamento de F\'{\i}sica,
Universidade Federal Rural do Rio de Janeiro\\
BR 465-07, 23890-971, Serop\'edica, Rio de Janeiro, Brazil\\
${}^{b}$Departamento de F\'{\i}sica, ICE, Universidade Federal de Juiz de Fora,\\
36036-330, Juiz de Fora, MG, Brazil\\
\bigskip
\today}


\begin{abstract}
\noindent It is a very well known fact that the energy conditions concerning traversable wormhole (WH) solutions of Einstein equations are violated.  Consequently, attempts to avoid the violation of the energy conditions constitutes one of the main areas of research in WH physics.  On the other hand, the current literature shows us that noncommutativity is one of the most promising candidates to help us to understand the physics of the early Universe.  However, since noncommutativity does not change the commutative results, we also can expect that energy conditions violation near the throat must occur.
We will show here that the violation of the energy conditions, described in a noncommutative space-time, has fixed conditions on the angular momentum of a rotating WH with constant angular velocity.   Also, we have established a new theoretical bound on the NC constant, $\theta$, as a function of some WH parameters.
\end{abstract}

\maketitle

\pagestyle{myheadings}
\markright{\it Rotating traversable wormholes in a noncommutative space.....}

\newpage



Dwelling in the imaginary of science fiction authors, a wormhole (WH) play a fundamental role when the issue is time travel.  Its popularity came from theoretical solutions of Einstein equations which show that WHs can act as tunnels linking two distant places \cite{basics} in the same (or not) Universe.  For review see \cite{visser,llo,basics2}.   Hypothetically, traversable WHs would permit the construction of time machines \cite{machine} and effective superluminal travels.  In this case however, the speed of light is not locally surpassed \cite{1}, of course.  In other words, WHs represent solution to Einstein's equations which are topologically non-simply connected.  Besides, they are interesting as compact objects that possibly exist in our Universe.

A traversable WH  permits (hypothetically) the passage of a traveler since the throat is not closed.  To be traversable, the WH should be finite and nonzero.  No curvature singularities and event horizon occur.
For this property to be accomplished it is considered that the WHs would be formed by a kind of {\it exotic matter} \cite{basics}, which violates the weak energy condition (WEC), and would violate the gravitational attraction (\cite{basics2} and references therein).  Therefore WHs have been viewed as quite different from black holes (BH).  Topologically speaking we can say that traversable WHs are short ``handles" or tunnels in spacetime topology that links, as we said above, widely separated regions of the same Universe.  Or  ``bridges," which links  two different spacetimes (different Universes).  In both cases we have non-trivial topologies ($R_1 \times R_2 \times S^2$) of multiply connected Universes.  Matter with negative energy density is well known in scenarios like the Casimir effect.

Although there is a quite large literature about WHs, we would like to describe very briefly some points that are important in this work.  To begin with, we know that to avoid singularities and horizons, one must construct the throat with nonzero energy-momentum tensor since the vicinity of the throat is the most important region of the WH.  In \cite{sh} the authors showed that under perturbation, there is a possibility that the WH collapse to a BH and the throat closes.  It is important to say that the process of WH creation, together with extremely large spacetime curvature can be both ruled by quantum gravity conditions \cite{basics2}.  Considering the process of WH formation, we assume that some field fluctuation collapsed and consequently it originates a rotating scalar field framework \cite{mn}.  This last one has one interior region, where the rotation is not zero, and two other exterior regions, one on each side of the throat, where the rotation is zero.  The internal rotation is supposed to keep the throat stable.  And the interior field is the source of the WH.  We will see that this rotation is present in the WH under the NC regime and it is sufficient to maintain the throat open.   Here we will analyze rotating WH with constant angular velocity.

There are various approaches dealing with WHs, both static \cite{2} and evolving relativistic forms \cite{3}.  Both consider static WH spacetimes sustained by a single fluid object that requires the violation of the null (NEC) and weak (WEC) energy conditions.  Violation of NEC on or near a traversable WH's throat is a general WH's property. The NEC is the weakest one.  Namely, if NEC is violated, all pointwise energy conditions will be violated too.  

Concerning the violation of NEC and WEC together with the property that keeps itself (the throat) open in order to permit traveling, we can say that the WHs must be formed by a (exotic)  matter, which is characterized by an energy-momentum tensor which violates the WEC.  Hence, the energy density must be negative in the framework of reference of at least a few observers.  Notice that although classical kinds of matter obey the WEC, quantum fields can generate negative energy densities (locally).  These last may be quite large at a given point.  

Therefore, to analyze the WH's energy conditions we have to verify the value of its energy-momentum tensor components.  After that, it was found that these traversable WHs have an energy-momentum tensor that violates NEC.  As a matter of fact, it was found that they violate all known pointwise energy conditions and averaged energy conditions.  Besides, they are very important to the singularity theorems and in classical BH thermodynamics.  

One of the main motivations to introduce NC features inside general relativity (GR) is the fact that noncommutativity can eliminate the divergences that dwell within GR \cite{6}.  It can be carried out by replacing point-like structures through noncommutativity and its smeared objects \cite{6.1}.  

Among all NC versions of GR there are some few which attack the WH subject although none of them discuss the energy conditions issues using canonical noncommutativity.  So, we believe that there is a gap concerning this energy subject in NC WHs (NCWHs) literature.   The other versions of NCWHs in the literature \cite{6.2} use Nicolini {\it et al} smeared objects approach \cite{6.3} which is based on coherent states, differently from us.  It is important to observe that we will consider noncommutativity via Bopp shift which was used by Chaichian {\it et al} in \cite{chaichian} in an operator context.

Since, as we said before, the traversability property of WHs are directed linked to the energy conditions and since its violation analysis is very well established in the literature, we deem that to show the conditions that the parameters of NCWHs have to obey in order to keep the energy conditions violated  at least at the Planck scale, can be a relevant result, since it is well known that noncommutativity does not modify the original physics of the system \cite{bbdp}.

Having said that, we will show that these energy violations constraints affect the angular momentum when the WH background geometry is NC.  Namely, the NCWH energy and flaring-out conditions are satisfied.  
In other words, we will introduce a quantum object, the NC parameter, inside the classical theory of WHs.  In this way we can say that we will construct a NC geometric background for WHs which do not obey the energy conditions.  The throat rotational property will be evident in this NC version of WH.  As we will demonstrate.  

We have chosen to analyze rotating WHs, which have angular momentum, of course, because it can be considered as the most general extension of the Morris-Thorne (MT) WH that one can consider \cite{teo}.   Besides, if $\omega$ is sufficiently large, we can have the presence of an ergoregion \cite{teo} where particles do not remain stationary with respect to infinity.   In fact, the magnitude of $\omega$ may have a small effect on the WEC violation considering a rotating WH \cite{kuhfittig}.  This conclusion is different from the case of a spherically symmetric WH, where a fast rotation originates a reduction in the case of WEC violation. 


\bigskip

We begin by writing the standard WH metric in the static, spherical and symmetric form with the coordinates $(t,r,\theta,\phi)$ as
\be
\label{1}
ds^2\,=\,-\,e^{2\Phi(r)}\,dt^2\,+\,\frac{1}{1-\frac{b(r)}{r}}dr^2\,+\,r^2\,(d\theta^2\,+\,sen^2\,\theta\,d\phi^2)
\ee
where $r\in (-\infty,+\infty)$ and we will assume that the redshift function $\Phi(r)=\;$constant to simplify our problem.  Eq. (\ref{1}) describes a WH geometry formed by two identical regions joined together at the throat.   However, we will consider in this paper the stationary and axially symmetric generalization of a MT \cite{basics}  WH, i.e., Eq. (\ref{1}), which would physically depict a rotating WH.   To be axially symmetric  means that the space-time possess a space-like Killing vector field $\partial/\partial \phi^a$, which generates invariant rotations concerning the angular coordinate $\phi$, namely, it is symmetric with respect to the axis of rotation.  
As described in \cite{teo,kuhfittig} the canonical metric for a stationary, aximmetric traversable  rotating WH is

\be
\label{1A}
ds^2\,=\,-\,N^2\,dt^2\,+\,\frac{1}{1-\frac{b(r)}{r}}dr^2\,+\,r^2\,K^2\,[\,d\theta^2\,+\,sen^2\theta\,(d\phi\,-\,\omega dt)^2\,]\,\,,
\ee

\ni where $N$ is the analog of the redshift function in Eq.(\ref{1}) and it ensures that there are no event horizon or curvature singularities; $b$ is the shape function which satisfies $b\leq r$ and obey the flare-out condition, at the throat, $b$ does not depend on $\theta$, namely, $\partial b / \partial \theta = 0$ at $r=r_0$; $K$ establishes the proper radial distance \cite{teo} measured at $(r, \theta)$ from the origin, i.e., $R=rK$, where $\partial R / \partial r > 0$ so that we can say that $2\pi kr sin \theta$ is the proper circumference of the circle at $(r, \theta)$. 
And $\omega$ is the angular velocity of the WH ($\omega = d \phi / dt$) relative to the asymptotic rest frame \cite{kuhfittig}.   In \cite{kuhfittig} it was obtained a general solution for the metric in (\ref{1A}) where it was assumed a time dependence of $\omega =\omega(r, \theta, t)$ and the WH cannot be considered as stationary.
All parameters in (\ref{1A}), i.e., $N, b, K$ and $\omega$, depend on $r$ and $\theta$.  Eq. (\ref{1A}) describes two identical, asymptotically flat regions linked at the throat $r=b>0$ which requires that $\omega \rightarrow \,0$ as $r \rightarrow \infty$.   We can, as always, imagine the surface as the connection between two asymptotically flat Universes, where $r$ decreases from $+\infty$ in the upper Universe to a minimum value $r=r_0$ at the throat.  And then it increases again to $+\infty$ in the lower Universe (\cite{kuhfittig} for details).
Hence, $r$ is asymptotically the proper radial distance.  In other words, if $$\omega
 = \frac{2L}{r^3}\,+\,O \Big(\frac{1}{r^4} \Big)\,\,,$$ then if we change to cartesian coordinates, we can show that $L$ is the total angular momentum of the WH \cite{teo}.   In the case of having mass and charge, the treatment is the standard one \cite{visser}.   It can be demonstrated \cite{teo,kuhfittig} that the WHs described by Eq. (\ref{1A}) also break the energy conditions.   The metric in (\ref{1A}) obviously reduces to the one in (\ref{1}), the MT one, in the limit of zero rotation and spherical symmetry \cite{kuhfittig}

\ba
\label{1AA}
N(r, \theta) \rightarrow e^{\Phi(r)}\,\,\:;\quad & & \qquad b(r, \theta) \rightarrow b(r)\,\,; \nonumber \\
K(r,\theta) \rightarrow 1 \qquad &\mbox{and}& \qquad \omega(r,\theta) \rightarrow 0\,\,,
\ea

\ni where we have the range $r \geq b$ and we can see en apparent singularity at $r=b \geq 0$ at the throat of the WH.  In particular, at the throat we will assume \cite{teo} that $N, K$ and $\omega$ are well behaved.  We would like to reinforce that here we will consider from now on that $\omega=\mbox{constant}$.

We will see that in the case of NCWH (WH described in a NC space-time), the breakdown of the energy conditions brings new conditions between the WH parameters described above.   The generalization of this formalism to the case of slow rotation WHs \cite{16C} is out of the scope of this work.

The Einstein field equations are given by $G_{\mu\nu}=8\pi T_{\mu\nu}$ and we will use that $c=G=1$ to obtain the following quantities
\be
\label{2} 
\rho(r)\,=\,\frac{b'}{8\pi r^2}
\ee
\be
\label{3}
p(r)\,=\,-\,\frac{1}{8\pi}\,\frac{b}{r^3} 
\ee
\be
\label{4}
p_{tr}\,=\,\frac{1}{8\pi}\(1-\frac{b}{r}\) \[ \frac{-b' r\,+\,b}{2r^2(r-b)}\] 
\ee
which are the energy density, the radial pressure and the transverse pressure respectively.  The shape function $b=b(r)$ is used to define the WH spatial shape, more specifically, the throat and $b'\,=\,db/dr$.  Using the metric singularity we have that
\be
\label{5}
\left. \(1\,-\,\frac{b(r)}{r} \) \right|_{r=r_0} \,=\, 0 \quad \Rightarrow \qquad b(r_{0})=r_0
\ee
where $r_0$ is the radius of the WH throat.  The condition for the space-time to be asymptotically flat is $b(r)/r \rightarrow 0$ as $|r| \rightarrow \infty$.  The shape function follows the conditions: $b'(r_{0}) < 1$ and $b(r) < r \,(r> r_0 )$.  The feature that the WH be traversable means that there are no event horizons or curvature singularities.  Hence, $\Phi$ (in (\ref{1}) and (\ref{1AA}))  must be finite everywhere.

Let us construct the shape function in a NC phase-space.  To carry out this task we will use the classical version of the idea of Chaichian {\it et al} \cite{chaichian} where the authors used the NC basic version of  \cite{li}

\be
\label{aaaaa}
\{\tilde{x}_i\,,\tilde{x}_j\}\,=\,\theta_{ij}\qquad,\qquad\{\tilde{x}_i\,,\tilde{p}_j\}\,=\,\delta_{ij}\qquad,\qquad\{\tilde{p}_i\,,\tilde{p}_j\}\,=\,0\,\,,
\ee
where the tilde denotes a NC variable and, of course, there is not any kind of operator in the metric in (\ref{1A}) or in any other metric.

To construct the NC version of the shape function we have to write, using (\ref{1}), that

\be
\label{6.1}
f(\tilde{r})\,=\,\left. \(1\,-\,\frac{b(r)}{r} \) \right|_{r=\tilde{r}}\,=\,0
\ee
where $\tilde{r}$ obeys the Poisson brackets given in (\ref{aaaaa}).

Let us now introduce the  WH in a NC space \cite{nasseri}

\be
ds^2\,=\,-\,N^2 dt^2\,+\,\frac{1}{1-\frac{b(\tilde{r})}{\sqrt{\tilde{r}\tilde{r}}}} d\tilde{r}^2\,+\,\tilde{r}^2 \,K^2\,[d\theta^2\,+\,sen^2\theta\,(d\phi\,-\,\omega dt)^2]\,\,,
\ee
where $\tilde{r}$ satisfies (\ref{aaaaa}).

It is {\bf very important} to understand that we are not considering any modification in Einstein's equations in this scenario \cite{setare}.   Obviously as the noncommutativity parameter is very small compared to the WH length scales, one can deal with the NC effects as some perturbations of the commutative counterpart, $f(r)$ in (7) written with the NC coordinates $\tilde{x}_i$ as

\be
\label{6.2}
1-\frac{b(\tilde{r})}{\sqrt{\tilde{r}\tilde{r}}}\,=\,0\,\,.
\ee

To recover space commutativity, as usual, we will introduce the Bopp shift
\be
\label{7}
x_i\,=\,\tilde{x}_i\,+\,\frac 12\,\theta_{ij}\,\hat{p}_j\,\,,
\ee
where $p_i = \tilde{p}_i$ and the un-tilde variables satisfy the usual Poisson brackets \cite{li},
\be
\label{6.1}
\{ x_i,x_j \} \,=\,0\quad,\quad \{ x_i,p_j \} \,=\,i \delta_{ij}\quad,\quad \{p_i,p_j \} \,=\,0\,\,.
\ee


From (\ref{6.2}) we can write that,
\ba
\label{9}
&&\frac{1}{\sqrt{\tilde{r}\tilde{r}}}\,=\,\frac{1}{\sqrt{\(x_i\,-\,\frac 12 \theta_{ij} p_j \)\(x_i\,-\,\frac 12 \theta_{ik}p_k \)}} \nonumber \\
&=&\frac{1}{r}\[1\,+\,\frac{1}{4r^2}\vec{L}\cdot\vec{\theta}\,-\,\frac{1}{8r^2}\(p^2 \theta^2\,-\,(\vec{p}\cdot\vec{\theta})^2 \) \] \,\,,
\ea
where $L_k\,=\,\epsilon_{ijk} x_i p_j\,,\,\,p^2=\vec{p}\cdot \vec{p}\,,\,\theta_k = \frac 12 \epsilon_{ijk} \theta_{jk}$, $\theta^2 = \vec{\theta} \cdot \vec{\theta}$ and we have used that $\epsilon_{ijl}\epsilon_{ikm}\,=\,\delta_{jk} \delta_{lm}\,-\,\delta_{jm} \delta_{lk}$.
We have performed an expansion in $\theta$ and we have used only first and second order terms in $\theta$ \cite{chaichian}.  Substituting Eq. (\ref{9}) in Eq. (\ref{6.2}) we have that
\be
\label{10}
b(r,p) \,=\,r\,-\,\frac{1}{4r}\vec{L}\cdot\vec{\theta}\,+\,\frac{1}{8r}\[p^2 \theta^2\,-\,(\vec{p}\cdot\vec{\theta})^2 \] \,\,.
\ee

Notice that we wrote $x_i \theta_{ij} p_j = \frac 12 \vec{L} \cdot \vec{\theta}$, where $\vec{L}$ can be understood as being connected to the throat's angular momentum \cite{teo,kuhfittig}.  For $\theta \rightarrow 0$ and $r \rightarrow r_0$ we have $b(r_0) \rightarrow r_0$, i.e., at the throat, as expected.  

For the equation of state we have that
\be
\label{11}
w(\tilde{r})\,=\,\frac{b(\tilde{r})}{\sqrt{\tilde{r}\tilde{r}}\,b'(\tilde{r})} \,\,,
\ee
which gives us the following result
\be
\label{12}
w(r)\,=\,\frac{1}{r^2}\left\{r^2\,-\,\frac{1}{4}\vec{L}\cdot\vec{\theta}\,+\,\frac{1}{8}\[p^2 \theta^2\,-\,(\vec{p}\cdot\vec{\theta})^2 \]\right\}\,\,.
\ee
This last equation is the equation of state for a NCWH.  Notice the presence of angular and linear momenta.  Differently from non-rotational NC black holes analyzed in \cite{nasseri}, here we will assume that these terms are connected with the WHs throat rotational features, as discussed before.


There are some alternatives to recover the obedience to the energy conditions. Recent cosmological observations has been shown that the cosmological fluid violates the strong energy condition (SEC)  and it implies, perhaps, that NEC can be possibly violated in a classical scenario.  Under these results one can conclude that WEC, NEC and the other energy conditions, cannot be considered underlying laws with consequences in the WHs construction.  However,  in \cite{3} it was shown that if we introduce a temporal function inside the WH metric, there will be constraints on this function that provide the WH to obey the energy conditions.  This situation is critical since it has been shown that classical systems violate all the energy conditions \cite{4}.   Besides, recent cosmological observations show us that possibly the cosmological fluid violates SEC and that NEC might possibly be violated in a classical behavior \cite{7}.  
Consequently, one of the main areas in WH investigation, as we said before, is to study how to avoid strongly the violation of NEC \cite{basics2} although it is well known that Morris and Throne \cite{basics} first formulated the {\it exotic matter} in terms of violations of WEC.
However, NEC and ANEC are always violated for WHs space-times.


Back to the disobedience to the energy conditions, our first step is to analyze the energy conditions so that, from Eqs. (\ref{2}) and (\ref{3}),
\ba
\label{13}
\tilde{\rho}\,+\,\tilde{p}\,&=&\,\frac{1}{\tilde{r}^2}\(b'\,-\,\frac{b}{\tilde{r}}\)\nonumber \\
&=&\frac{1}{4 r^4}\left\{\vec{L}\cdot\vec{\theta}\,-\,\frac{1}{2}\[p^2 \theta^2\,-\,(\vec{p}\cdot\vec{\theta})^2 \] \right\} \nonumber \\
&\times&\[1\,+\,\frac{1}{2r^2}\(\,\vec{L}\cdot\vec{\theta}\,-\,\frac{1}{2}\[p^2 \theta^2\,-\,(\vec{p}\cdot\vec{\theta})^2 \] \) \] \\
&=&\frac{1}{8r^6}\, \left\{(\vec{L} \cdot \vec{\theta})^2\,+\, (\vec{L} \cdot \vec{\theta})2r^2\,-\,r^2 \[p^2 \theta^2\,-\,(\vec{p}\cdot\vec{\theta})^2 \] \right\} \quad \leq\:\:0 \nonumber
\ea

\ni where we have considered only first and second order terms in $\theta$, which is usual in NC literature and $L=\mbox{constant}$. 

Since $r^6 > 0$, let us rearrange Eq. (\ref{13}) in a way that

\be
\label{AA}
(\vec{L} \cdot \vec{\theta})^2\,+\, (\vec{L} \cdot \vec{\theta})2r^2\,-\,r^2 B^2\,\leq\,0\,\,,
\ee

\ni where $B^2 = p^2 \theta^2\,-\,(\vec{p}\cdot\vec{\theta})^2$, which can be manipulated considering only $\theta_3 = \theta$ and that the other $\theta$ components are zero.  Consequently, we also have that $\vec{L} \cdot \vec{\theta} = L_z \theta = L \,\theta$ and $\vec{p} \cdot \vec{\theta} = p_z \theta$.   In this way we can write that

\ba
\label{AA1}
B^2 &=& p^2 \theta^2\,-\,(\vec{p}\cdot\vec{\theta})^2  \nonumber \\
&=&(\,p_x^2\,+\,p_y^2\,+\,p_z^2\,)\,\theta^2\,-\,(p_z \theta)^2 \nonumber \\
&=&M^2\,(\,v_x^2\,+\,v_y^2 \,)\, \theta^2\,\,,
\ea

\ni where $M$ is the exotic mass of the WH and the maximum value of $v^2 = 1$.  So, equation (\ref{AA1}) can be written as

\be
\label{AA2}
B^2\,=\,M^2\, \theta^2\,\,.
\ee

Back to Eq. (\ref{AA}) we have that

\be
\label{AA3}
L^2 \theta^2\,+\,2 r^2 L\, \theta \,-\,r^2 M^2 \theta^2 \leq 0
\ee

\ni and solving this inequality for $L\,\theta$ we have that

\ba
\label{AA4}
-r^2\,-\,r^2 \sqrt{1\,+\,\frac{M^2 \theta^2}{r^2}} &\leq& L \theta \leq -r^2\,+\,r^2 \sqrt{1\,+\,\frac{M^2 \theta^2}{r^2}} \nonumber \\
\Longrightarrow \quad \mbox{at the throat} \Longrightarrow -2r^2_0\,-\,\frac{B^2}{2} &\leq& L \theta \leq \frac{M^2 \theta^2}{2}\,\,,
\ea

\ni where we can use the bound for $\theta$ as being $\theta \geq 4.35\,l_P^2$ \cite{nasseri}, to eliminate the square root, since $\frac{M^2 \theta^2}{r^2_0}\ll 1$.  Considering that $L > 0$, we can rewrite Eq. (\ref{AA4}) as

\be
\label{AA5}
0 \leq L \leq \frac 12 M^2 \theta
\ee

Now let us carry out the summation $\tilde{\rho}\,+\,\tilde{p}_{tr}$ which is given by
\ba
\label{14}
&&\tilde{\rho}+\tilde{p}_{tr}\,\,  \nonumber \\
&=&\frac{1}{r^2}\,
\[1\,+\,\frac{1}{4r^2}\,\ \vec{L}\cdot\vec{\theta}\,-\,\frac{1}{8r^2}\[p^2 \theta^2\,-\,(\vec{p}\cdot\vec{\theta})^2 \] \]  \nonumber \\
&\times& \left\{ 1\,+\,\frac{1}{2r^2}\,\vec{L}\cdot\vec{\theta}\,-\,\frac{1}{4r^2}\[p^2 \theta^2\,-\,(\vec{p}\cdot\vec{\theta})^2 \]  \right\}  \\ 
&=&\frac{1}{8r^6}\,\Big[ (\,\vec{L} \cdot \vec{\theta}\,)^2\,+\, 6r^2\, \vec{L} \cdot \vec{\theta}\,-\,3r^2 B^2\,+\,8r^4 \Big] \quad \leq \:\:0 \nonumber\,\,,  
\ea

\ni and analogously, we can  rewrite (\ref{14}) as

\be
\label{BB}
\frac{1}{8r^6}\,\Big[ (\,L \,\theta\,)^2\,+\, 6r^2\, L\, \theta\,-\,3r^2 M^2 \theta^2\,+\,8r^4 \Big] \quad \leq \:\:0
\ee

\ni and the solution at the throat is

\be
\label{BB1}
0 \leq L \leq \frac 32 M^2 \theta \,-\,\frac{2r_0^2}{\theta}\,\,.
\ee

\ni If we use the bound $L > 0$ we have that the right side of (\ref{BB1}) must be positive, hence

\be
\label{BB2}
\frac 34 \frac{M^2 \theta^2}{r_0^2} > 1
\ee

\ni at the throat of the WH.   From  (\ref{BB2}) we can establish a new theoretical bound for $\theta$

\be
\label{BB3}
\theta^2 > \frac 43 \frac{r_0^2}{M^2}\qquad \Longrightarrow \qquad\theta < \frac{2}{\sqrt{3}} \frac{r_0}{M}\,\,.
\ee

\ni Comparing (\ref{AA5}) and (\ref{BB1}) and using (\ref{BB3}) we conclude that

\be
\label{BB4}
L \leq \frac 12 M^2 \theta\,\qquad \Longrightarrow \qquad L <\,\frac{1}{\sqrt{3}}\,r_0\,M \,\,,
\ee

\ni which is a new bound for the angular momentum of our rigid rotating WH, which could be established thanks to the NC framework of space-time.


\bigskip
\bigskip


To conclude this work, the analysis of a NC geometrical description of a rigid rotating WH could reveal some informations concerning the WH parameters.  In the case discussed here, where we have considered a rotating WH with constant angular velocity and constant angular momentum at the throat, we have obtained new bounds for the angular momentum and for the NC parameter $\theta$ as functions of $r_0$, the radius of the throat, and $M$, the mass of the exotic matter.

As future perspectives, since we can consider that the solution of Einstein's equations permit us to elaborate WH physics, a natural way is to add a cosmological constant to the analysis of NCWH framework obtained here.  Hence, in the light of noncommutativity \cite{jhep-nosso} we can ask how to deal with a NCWH formulation of the cosmological constant.  Or if theories that yields bouncing cosmological models \cite{poulis} also generate NCWH solutions.  We can ask also what can be revealed when a NCWH collapses since its commutative partner collapses to become a BH.

In this work we have introduced noncommutativity through the Bopp shift.  However, we can imagine the construction of a WH inside a totally NC spacetime, like the one developed in \cite{amorim,abreu} which is known as the extended Doplicher-Fredenhagen-Roberts space-time.  We may ask what would be the WH ten dimensional structure within a space-time where $\theta^{\mu\nu}$ is a coordinate with a conjugate momentum instead of a simple NC parameter, as in the majority NC literature.  Another way to introduce noncommutativity in WH theory is through its Lagrangian formulation.  A comparison can be accomplished between both methods.

It is well known that even below the Planck scale, there exists the possibility of other classical violations of NEC, such as higher derivative models, Brans-Dicke theory, or else \cite{bv}.  But all these last ones are based on modifications of general relativity at high energies.
Facing the results obtained here, we believe that an investigation of these issues in a NC geometry background would lead us to interesting paths not explored so far.

\bigskip
\bigskip

\noindent EMCA would like to thank CNPq (Conselho Nacional de Desenvolvimento Cient\' ifico e Tecnol\'ogico), a Brazilian research support agency, for partial financial support.

\end{document}